# The mesospheric inversion layer and sprites

S. Fadnavis[1], Devendraa Siingh[1*], R.P. Singh[2]

[1] Indian Institute of Tropical Meteorology, Pune-411 008, India.
[2] Vice-Chancellor, Veer Kunwar Singh University, Ara (Bihar)-802301, India.
* also at, Center for Sun-Climate Research, Danish National Space Center, DK-2100, Copenhagen, Denmark

**Abstract**

The vertical structure of temperature observed by SABER (Sounding of Atmosphere using Broadband Emission Radiometry) aboard TIMED (Thermosphere, Ionosphere, Mesosphere Energetics and Dynamics) and sprites observations made during the Eurosprite 2003-2007 observational campaign were analyzed. Sprite observations were made at two locations in France, namely Puy de Dôme (45° 46' 19.2''N; 02° 57' 44.64''E; 1.464 km altitude) in the French Massif Central and at the Pic du Midi (42$^0$ 56' 11''N; 00$^0$ 08' 34''E; 2.877 km altitude) in the French Pyrénées. It is observed that the vertical structure of temperature shows evidence for a Mesospheric Inversion Layer (MIL) on those days on which sprites were observed. A few events are also reported in which sprites were not recorded, although there is evidence of a MIL in the vertical structure of the temperature. It is proposed that breaking gravity waves produced by convective thunderstorms facilitate the production of (a) sprites by modulating the neutral air-density and (b) MILs via the deposition of energy. The same proposition has been used to explain observations of lightings as well as both MILs and lightning arising out of deep convections.

**1. Introduction**

Mesospheric Inversion Layers (MILs) over low and mid-latitudes have been reported in a number of studies [*Schmidlin,* 1976; *Hauchecorne et al.,* 1987; *Leblanc and Hauchecorne,* 1997; *Meriwether et al.,* 1998; *States and Gardner,* 1998; *Kumar et al.,* 2001; *Fadnavis and Beig,* 2004; *Fadnavis et al.,* 2007]. There are two types of MILs (1) lower MILs (~75km) and (2) upper MILs (~90 km) [*Fadnavis and Beig,* 2004; *Meriwether and Gerrard,* 2004]. Possible mechanisms for production of MILs have been reviewed by [*Meriwether and Gardner,* 2000]. *Meriwether and Gerrard* [2004] stated that the lower MILs are the result of dissipating planetary waves and a mechanism responsible for the formation of the upper MIL may be nonlinear gravity wave-tidal interactions. In some studies [*States and Gardner, 1998; Chu et al.,* 2005] the upper MIL is attributed to the diurnal tides, whereas others [*Meriwether et al.,* 1998; *Liu and Hagan,* 1998; *Liu et al.,* 2000; *Oberheide et al.,* 2006] proposed the coupling of gravity waves (GWs) and the mesopause tidal structures to produce the upper MILs. *Thomas et al.* [1996], *Gardner and Yang* [1998] and *Meriwether et al.* [1998] indicated that GW breaking plays an important role in the development of the upper MIL. Two-dimensional modeling studies reveal that GWs play a significant role in increasing the temperature amplitude of the tidal structure and thus contribute to MIL [*Hauchecorne and Maillard,* 1990; *Leblanc et al.,* 1995].

Gravity waves are produced by oscillating updrafts and downdrafts in convective systems [*Fovell and Ogura,* 1989]. The excited gravity wave amplitudes are proportional to the convective core updraft velocity, stronger updrafts lead to large-amplitude-gravity waves. Gravity waves propagating upward into the high atmosphere may amplify and become a major source of mechanical energy and momentum into this region from below [*Swenson and Liu,* 1998]. Breaking gravity waves modulate neutral density in the mesosphere: reduced density values will facilitate the electrical breakdown process [*Sao Sabbas et al.,* 2009]. This concept has been used to explain modulated optical emissions in the nightglow layer, which are detectable from the ground [*Taylor and Hill,* 1991; *Turnbull and Lowe,* 1991] as well as from orbiting imaging platforms [*Dewan et al.,* 1998; *Picard et al.,* 1998]. *Sentman et al.* [2003] documented the first complete ring structure of OH emissions modulated by thunderstorms. Their analysis revealed that short period concentric wave structures emanated radially outward from a central excitation region directly above the storm whose period gradually increases and wave length decreases with time.

Sprites are luminous events observed in the mesosphere in association with lightning ground flashes of positive polarity [*Boccippio et al.,* 1995; *Singh et al., 2004; Siingh et al., 2005; Siingh et al.*, 2007 *with reference therein*]. Several mechanisms based on the relationship between positive electrical charges in the troposphere and a physical process in the mesosphere is proposed for the production of sprites. The physical processes include sprite production by electrical breakdown [*Wilson,* 1925; *Huang et al.* 1999], electrical heating [*Pasko et al.,* 1995], electron runaway [*Bell et. al.,* 1995], runaway electron beams [*Roussel-Dupre and Gurevich,* 1996; *Lehtinen et al.,* 1997], high altitude gamma rays [*Taranenko and Roussel-Dupre,* 1996] and infrared glow from $CO_2$ emission [*Picard et al.,* 1997]. A few cases of sprites induced by negative cloud-to-ground flashes have also been reported [*Barrington-Leigh et. al.,* 1999] as discussed in *Williams et al.* [2007], who investigated the origins of the pronounced polarity asymmetry for sprites. *Williams* [1998] reviewed the atmospheric electrical and meteorological observations and proposed that sprites are produced by laterally extensive mesoscale convective systems (MCSs) in which the positive charge reservoir predominates in the 4-6 km of altitude. MCSs are characterized by laterally extensive regions of stratiform precipitation with a total area larger than $10^4$ km$^2$, which is usually more than an order of magnitude greater than the area of ordinary thunderstorms. It is also suggested that sprites are associated with "spider lightning" which is extended horizontal lightning in clouds that may feed several CG flashes [*Mazur et al.,* 1998]. Recently *Soula et al.* [2009a] analyzed two storms, each producing 27 sprite events observed during the EuroSprite campaigns. Both storms were identified as MCSs with a trailing stratiform configuration and reaching a maximum cloud area of $\sim 12 \times 10^4$ km$^2$.

Sprites may affect the ionization of the mesosphere which can be estimated from the observations of amplitude perturbations of VLF waves propagating through it. Based on EuroSprite-2003 observations, *Mika et al.* [2005] had suggested that sprites are nearly always accompanied by "early" VLF perturbations having long recovery times (~30–300s); which suggests spatially extended, diffuse regions of electron density increases at altitudes higher than 75 km [*Neubert et al.,* 2008]. This supports theoretical predictions of air breakdown in the upper D-region ionosphere during sprite occurrences, triggered by strong quasi-electrostatic (QE) fields [*Wilson,* 1925]. Another category of VLF perturbations with a long onset duration of up to 2.5s was observed in relation to ~55% of the sprite events as discussed by *Neubert et al.* [2008]. They have suggested that the long onset duration is a result of secondary ionization build-up in and/or heating of the upper D-region below the nighttime VLF reflection height, caused by electromagnetic pulses radiated upwards from horizontal intra-cloud discharges. The radiation fields accelerate sprite-produced electrons which then impact to produce secondary electrons in an avalanche process leading to ionization build-up [*Haldoupis et al.,* 2006]. The enhancement in electrical conductivity of mesosphere during sprite was supported by *Williams et al.* [2006] based on "glow discharge tube" measurements, who observed electrical conductivity $\sim 3 \times 10^{-5}$ S/m for a radiance value in the range of sprite measurement. This value is some three orders of magnitude greater than the ambient electrical conductivity at the simulated altitude (~65km).

Sprites may also produce NO$_x$ in the stratosphere [*Sentmen et al.,* 2008; *Enell et al.,* 2008; *Neubert et al.,* 2008; *Siingh et al.,* 2008]. The production may involve excitation of N$_2$ into short-lived states which cascade into the metastable $N_2$ ($A^3\Sigma_u^+$) state, ionization of $N_2$ and $O_2$, and dissociation of $O_2$ [*Enell et al.,* 2008]. In regard to $NO_x$ production, the *Sentman et al.* [2008] study showed that the principal reactions leading to production of $NO_x$ were dissociation of $N_2$ in the head of a sprite-streamer leading to creation of active atomic nitrogen N (2D), which then interacts with $O_2$ to produce NO. *Enell et.al.* [2008] estimated the total production of $NO_x$ to be around five times the background in the streamers, whereas the model of *Gorodillio-Vajquez* [2008] predicts NO and $NO_x$ enhancements of a factor of 10. *Neubert et al.* [2008] estimated the total number of NO molecules produced in a streamer ~$1.5\times10^{19}$, with an estimated average total production on the order of $10^{23}$–$10^{25}$ molecules per streamer. Considering the global occurrence rate of sprites to be 3/min., the total global production of $NO_x$ ~$10^{31}$ molecules per year [*Neubert et al.,* 2008; *Sentman et al.,* 2008], which on global scale is quite small [*Neubert et al.,* 2008, *Siingh et al.,* 2008]. The local enhancements of sprite-induced $NO_x$ increases with altitude from a few percent at 47 km to tens of percent at 60 km [*Neubert et al.,* 2008], thus being roughly consistent with Sodankyla Coupled Ion-neutral Chemistry (SIC) model estimates [*Enell et al.,* 2008]. Data from the Global Ozone Monitoring by Occultation of Stars (GOMOS) instrument on ENVIronmental SATellite (ENVISAT) did not support enhancements on larger regional scales [*Rodger et al.,* 2008]. The enhancement in $NO_x$ concentration decreases ozone concentration through catalytic cycles. The impact of the sprite-produced NO on ozone was estimated to decrease the ozone concentration by a few percent in the 50-60 km altitude range [*Enell et al.,* 2008]. Thus, significant ozone perturbations from sprites are unlikely, except for sprite-active storms.

Recently *Fadnavis et al.* [2007] have analyzed the vertical profiles of temperature, the mesospheric ozone volume mixing ratio and the thunderstorm occurrence rate for the same period and studied the possible relationships between thunderstorms and the mesospheric heating near 70-85 km altitude. They showed a fairly good correlation (correlation coefficient lies between 0.50 and 0.72) between the inversion days (days on which MIL was observed in temperature profile) and thunderstorm activity. They have also reported some inversion days on which thunderstorms were not recorded at the corresponding station. The seasonal variation of frequency of the occurrence and the amplitude of the inversion also exhibits a fairly good correlation (correlation coefficient > 0.5) with the seasonal variation of ozone concentration over these regions. *Fadnavis et al.* [2007] considered that gravity waves produced during convective thunderstorms propagating upwards become unstable in the mesosphere and that gravity waves break there [*Thomas et al.,* 1996]. The turbulent heating, arising from the breaking of waves, provides a feedback mechanism that then may maintain the observed MIL [*Wu and Killeen,* 1996; *Fadnavis et al.,* 2007]. Exothermic chemical reactions involving ozone also produce heating [*Mlynczak and Solomn,* 1993], which may add to gravity wave heating. This may explain the positive correlation between thunderstorms and MILs.

Figure 1 shows the location of blue jets, red sprites and elves with height. In the same figure, the variation of temperature and occurrence of mesospheric inversion along; with the demarcation of the troposphere, stratosphere, mesosphere and thermosphere are also given [*Lyons et al.,* 2000, *Neubert,* 2003, modified figure]. Sprites and the mesospheric inversion layer occur at the same height in the atmosphere. An attempt is made to understand the relationship between sprites and MIL through a physical mechanism involving gravity waves.

**2. Data and Analysis**

The Sounding of the Atmosphere using Broadband Emission Radiometry (SABER) instrument is one of four instruments mounted on the Thermosphere Ionosphere Mesosphere Energetics and Dynamics (TIMED) satellite [*Christensen et al.,* 2004]. It was launched in December 2001 into a 625 km orbit inclined at 74 degrees to the equator. The TIMED satellite orbits the Earth about 15 times per day, and measures approximately 100 vertical profiles of temperature on each orbit. SABER has a vertical resolution of about 2 km in the altitude range 10–105 km, and an along-track resolution of approximately 400 km. It is used to study the mesosphere and lower thermosphere/ionosphere region of the Earth's atmosphere. It is a 10-channel infrared (1.27 to 17 µm) radiometer that observes atmospheric emissions from the Earth's limb. The collected data are ground-processed to retrieve temperature, ozone, water vapor, carbon dioxide, and key parameters describing the energetics of the high atmosphere. Kinetic temperatures are determined from 15μm $CO_2$ emissions, taking account of the non-LTE (non local thermodynamic equilibrium) conditions in the MLT region, in combination with 4.3μm $CO_2$ emission used to derive the $CO_2$ volume mixing ratio [*Mertens et al.,* 2004]. Preliminary estimates of temperature error are quoted as 5K for the systematic error and 2K for the accuracy at 87 km.

The temperature measurements are continuously available from January 2002 at the website: http://saber.gats-inc.com/index.php. The SABER (Level 2A) temperature results for the period 2003 to 2007 are used for this study. The daily temperature profiles from 50 km to 100 km show the most notable feature "inversion" in the temperature structure of mesosphere. The days on which the vertical profile of temperature shows a MIL are known as inversion days and the other days are known as non-inversion days (Figure 2). The MILs of amplitude 10 K or more can be considered as significant [*Meriwether and Gerrard,* 2004, *Fadnavis and Beig* 2004; *Fadnavis et al.,* 2007]. The MILs are found as a layer ~10-15 km vertical thickness within the upper mesosphere at the equatorial and mid-latitude regions [*Meriwether and Gerrard,* 2004]. Hence small perturbations (near 80 km) with rise in temperature less than 10K and vertical thickness less than 10km in vertical profile of temperature, as seen in figure 3 are not considered as MILs.

To understand the relationship between mesospheric inversion layers and sprites, observations of sprites during the EuroSprite (2003-2007) campaign were analyzed along with the SABER temperature profiles. The sprite observations were made at two locations in France, namely Puy de Dôme (45° 46' 19.2''N; 02° 57' 44.64''E; 1.464 km altitude) (in the French Massif Central) and Pic du Midi ($42^0$ 56' 11''N; $00^0$ 08' 34''E; 2.877 km altitude) (in the French Pyrénées) [*Chanrion et al.,* 2007; *Neubert et al.,* 2008]. Sprites were detected for the period 2003-2007 in the months of July-August-September-October. Transient luminous events were detected above MCSs using ground-based observing facilities during four years "Coupling of Atmospheric Layers (CAL)" observations [*Neubert et al.,* 2008]. Vertical temperature profiles obtained from SABER instrument, approximately above (20 to 140 km altitude and $3^0 \times 3^0$ grid) the Puy de Dôme and Pic du Midi locations are analyzed.

During the Eurosprite campaigns thunderstorms were monitored with a more complete set of instrumentation, thereby allowing a better characterization of cloud and lightning properties of the sprite-producing regions of the storms. The details about the experiments of these stations and other information with camera details have been discussed by *Chanrion et al.* [2007], *Neubert et al.* [2008] and *Sao Sabbas et al.* [2009]. Short range lightning detection system SAFIR were located in northern Spain and in the Provence region of Southern France. Each has three VHF ( ~ 100 MHz) receivers in a triangle with ~100 km distance between the receivers that measure activity in clouds, giving information on the location (in the horizontal plane) and time of IC (intra-cloud) and CG (cloud-to-ground) flashes (*Drüe et al., 2007*). Additional information came from a broad-band, VLF (~ 10 kHz) receiver located at Nancay. CG activity was also available from conventional lightning detection networks provided by the French Météorage system of 17 sensors covering the whole country. Each sensor uses two techniques for CG location by direction finding (DF) and time of arrival (TOA). The characteristics of the CG flashes include the location (horizontal), the time of occurrence, the polarity, the multiplicity and the peak currents of each stroke [*Cummins et al.* 1998]. Other characteristics of thunderstorms are derived from meteorological radars. A network of weather radars, ARAMIS, run by MeteoFrance, covers France. Eighteen radars (ten C-band and eight S-band) with 250 km range provide complete horizontal scan every five minutes, and provide the reflectivity factor which is directly related to the precipitation rate. Sprites can then be related to the cloud system properties, to determine the type of thundercloud structure that favours their production and in what evolutionary phase of the convective system they are generated [*Neubert et al., 2008; Sao Sabbas et al., 2009; Soula et al., 2009a*].

**3. Results and Discussion**

Lightning data collected at Pic du Midi ($42^0\ 56^{'}\ 11^{''}$N; $00^0\ 08^{'}\ 34^{''}$E; 2.877 km altitude) in the French Pyrénées during the EuroSprites (2003-2007) campaign are analyzed to understand relationship between lightning and sprites. It is noted that there are days with convective clouds but neither lightning nor sprites are observed. It is known that the vast majority of cloud systems having some space charge do not meet the requirements for lightning. Convective clouds, with vertical motions of at least 5 to 10 ms$^{-1}$ and a substantial depth of mixed phase precipitation ($-10^{o}$C to $-40^{o}$C) are required for most lightning (*Lyons, 2006*). Table 1 presents randomly selected days when lightning in Europe are recorded but sprites are not observed. In figure (4), we have shown number of lightning flashes per ten minutes on the August 13, 2005 and July 01, 2006. Large numbers of flashes suggest thunderstorms to be deep convective system. Thus deep convective system having lightning flashes may not initiate sprites. In this table there are some days on which MIL has been observed and on the remaining days MIL has not been observed. Table 2 presents randomly selected days when both lightning and sprites are observed. Almost on all these dates evidence of MIL is also observed. *Neubert et al.* (2008) analyzed the storms of July 21, 2003; July 29, 2005; September 11, 2006; and many other storms in order to separate convective and stratiform precipitations. They have reported that almost all storms begin small with predominantly intense convective precipitation (>50 dBZ) with a trailing stratiform configuration. Sprites were observed during periods where the area of the stratiform precipitation region with reflectivities from 30-45 dBZ was growing, and usually lasted until the size of this region begins to decrease. For the storm systems observed during the EuroSprite campaigns, sprites tended to occur when the cloud canopy reached its largest extent [*Neubert et al.,* 2008 ; *Soula et al.,* 2009a]. In all these cases charge moment change ($\Delta M_q$) is large (~800 C km), which is the key source parameter in the initiation of sprites [*Wilson,* 1925, *Huang et. al.,* 1999; *Lyons* 2006]. These discussions show that there is not one to one correspondence between convective cloud discharges and sprites. Only certain classes of convective storms can generate cloud – to - ground discharges with the requisite large $\Delta M_q$, which could initiate sprites.

We have also computed the fraction of sprite days with respect to lightning days. The analysis of data shows observations of sprites on 10 days and lightning on 69 days in the year 2003. Thus, the ratio of sprite days to lightning days is about 0.144. In the year 2005, the days with sprites and lightning are 12 and 142 respectively, yielding the ratio to be about 0.084. The days with sprites and lightning in the year 2006 are 7 and 108 respectively. The ratio is about 0.064. In 2007, the days with sprites and lightning are 8 and 132 respectively and the ratio is about 0.06. The average ratio for these years comes out to be about 0.088.

To understand the relation between the MIL and sprites occurring above the thunderstorms, temperature profiles were plotted on each of the days on which sprites are observed above the thunderstorms (*Neubert and Chanrion, 2008; www.electricstorms.net)*. Figures 2 (a) to (f) exhibit typical vertical temperature structure for randomly selected

246 inversion days 22-07-2003 (near Pic), 27-07-2003 (near Puy), 11-08-2005 (near Puy), 09-09-
247 2005 (near Puy), 27-07-2006 (near Puy), and 13-10-2007 (near Pic) respectively. It is quite
248 evident that on an inversion day the temperature increases by ~20K at about 100 km altitude.
249 At these stations the inversions occur mostly between 75-95 km.
250 Table 3 depicts the days during the campaign on which sprites are observed above the
251 stratiform region of convective thunderstorms and the respective vertical profile of
252 temperature shows the evidence of MIL on the same day. The brightest region of the sprites
253 lies in the altitude range 65-75 km, above which there is often faint red glow or wispy
254 structure that extended to about 90 km (Figure 1). Below the bright red region, tendril-like
255 filamentary structures often extend downward to as low as 40 km. Figure 3 depicts the
256 vertical temperature structure for the randomly selected typical non-inversion days (a) 25-07-
257 2003 (near Pic), (b) 26-08-2003 (near Puy), (c) 18-07-2005 (near Puy), (d) 20-07-2005 near
258 Puy, (e) 06-07-2006 (near Puy), (f) 14-10-2007 (near Pic) as recorded by SABER near Pic du
259 Midi or Puy de Dôme where simultaneously sprites are not observed on the same day at these

260 locations. Table 4 depicts the days on which sprites are not recorded during the campaign and
261 the vertical profile of temperature does not show the existence of inversion (non–inversion
262 days). In figure 5 we have presented six cases of MIL which were not associated with sprites.
263 These are explained by the fact that MILs are generated by the interaction of gravity waves
264 generated during thunderstorms with planetary waves, as well as contributions from others
265 sources [*Meriwether and Gardner, 2000*].
266     Comparing tables 1 and 4, it is noted that at least there are six days on which lightning
267 discharges have been observed but sprites and MIL have not been observed. It has already
268 been discussed that sprite-initiation requires large value of $\Delta M_q$ and hence thunderstorms on
269 those dates may not have sufficient value of $\Delta M_q$ and other required properties. Gravity
270 waves play an important role in the development of MIL [*Meriwether et al*., 1998; *Gardner*

and Yang, 1998]. The absence of MIL could be explained by considering the fact that either gravity waves are not generated during the thunderstorm activity period or generated gravity waves could not reach the mesosphere due to filtering effect of the intervening medium. Even gravity waves reaching the mesosphere with small amplitudes may not be able to produce observable MIL effect.

Convective motion associated with thunderstorms generate gravity waves which propagate upwards and may get amplified [*Sentman et al., 2003; Lyons and Armstrong, 2004*]. The gravity waves become unstable at the height where the zonal wind velocity becomes equal to the wave phase velocity. When this occurs, convective instability or shear instability causes transfer of wave energy from gravity waves to the mean flow with consequent changes in the tidal wind amplitude and phase. Usually in the mesosphere this condition is satisfied and breaking of gravity waves takes place [*Sica and Thorsley, 1996: Thomas et al., 1996*]. The turbulent heating, arising from the breaking of waves, provides a feedback mechanism that then may maintain the observed MIL [*Fadnavis et al., 2007*]. The plausible role of breaking gravity waves in the formation of the mid-latitude temperature inversion was also pointed out by *Leblanc and Hauchecorne* [1997] and *Williams et al.* [2002]. Based on the experimental and simulation results *Oberheide et al.* [2006] have suggested that the interaction of gravity waves with planetary waves results in a sharp phase transition of planetary waves large enough to create a MIL.

Convective thunderstorm generated gravity waves [*Dewan et al., 1998*] while propagating upward produce perturbations in the neutral density, typically at scales of 10-50 km and $\Delta N/N \sim 30\%$ (where N is neutral density), as evidenced in images of OH airglow. These perturbations could modulate the local threshold electric discharge field, with low densities having lower thresholds [*Sentman et al.,* 2003; *Sao Sabbas et al.,* 2009]. For a given charge moment change in the underlying thunderstorm the favoured initiation point for sprites would be where threshold electric field is minimum [*Sao Sabbes et al.*, 2009]. *Soula et al.* [2009b] have presented some structure sprites were linked with the electrical perturbation of the elves (6 elves (4 with a sprite)) and halos (5 halos (4 with a sprite)) observed at the night of 15/16 November, 2007 during the coordinated European campaign of TLE observation in 2007 over the Mediterranean Sea. *Neubert et al.* [2005] had presented a structured sprite cluster that was explained to have been formed in such a process. However, it is not known whether the medium scale structure (~10 km) is caused by the local conditions such as the turbulence field in the mesosphere or by interference patterns in the radiated and reflected source electric fields [*Neubert et al.*, 2005]. The electric fields from upward discharges [*Krehbiel et al.*, 2008] decrease as the inverse of the square of distance whereas the threshold for electrical breakdown follows the density of air which decreases exponentially. Under these circumstances there exists an altitude range, well above thunderstorm, where the electric field can exceed the threshold for breakdown and initiate sprites [*Wilson*, 1925; *Huang et al.*, 1999; *Williams*, 2001].

The sprites can change the chemical composition of the atmosphere by producing $NO_x$ and $HO_x$ in the mesosphere and lower atmosphere [*Sentman et al.*, 2008]. Observations show that $N_2$ is ionized during sprites [*Armstrong et al.*, 1998; *Heavener*, 2000]. Ionization in the mesosphere is primarily in the form of positive and negative ions [*Brasseur and Solomon*, 1986] and because of their low mobility and the small electric field they would remain where they are created. The ions help in the production of $NO_x$ and $HO_x$. *Enell et al.* [2008] and S*entman et al.* [2008] have estimated the chemical effect of sprites in the atmosphere and suggested that sprites are of little global chemical significance in producing $NO_x$ as compared to the other sources. This is supported by the measurements of $NO_2$ at 52 km attitude over large thunderstorms which showed enhancements by ~ 10% [*Arnone et al.*, 2008]. However in extraordinary cases local enhancements of $NO_x$ may became significant, up to 5 times the minimum background at 70 km. The enhancement in $NO_x$ causes reduction in ozone in the mesosphere [*Singh et al.,* 2005; *Schumann and Huntrieser,* 2007; *Sentman et al.,* 2008; *Siingh et al.,* 2008]. *Fadnavis et al.* [2007] observed that the ozone concentration is halved within the MIL.

*Sentman et al.* [2003] investigated simultaneous observations of coincident gravity waves and sprites to establish an upper limit on sprite-associated thermal energy deposition in the mesosphere. They estimated the resultant temperature perturbation associated with the concentrically expanding gravity waves to be ~7K. The neutral temperature perturbation $\Delta T$ caused by sprites $\leq 0.5K$ at the altitude ~85 km. The corresponding total thermal energy deposited by the sprite is less than ~ 1 GJ [*Sentman et al.*, 2003]. The chemical changes by sprites are fairly minor, and the heating is minuscule, and they could not be responsible in producing the observed MIL. This suggests that sprites may not contribute in the production/ maintenance of MILs, whereas the density reductions that accompany dissipating gravity waves may help in the initiation of sprites.

**4**. **Conclusions**

By analyzing radio soundings of the mesosphere (vertical temperature profile) and optical data from the EuroSprites campaign during 2003-2007, we have shown an association of sprites with the inversion in the mesospheric temperature profile, known as MIL. Gravity waves generated during convective thunderstorms propagate upwards into the mesosphere; they facilitate the production of MIL by depositing mechanical energy and momentum into this region from below. Gravity waves also modulate the neutral air density in the mesosphere. The intense electric fields caused by the charge imbalance in the stratiform regions of convective thunderstorm after a CG flash penetrates into the mesosphere and initiate the production of sprites. Modulation of the neutral air density and heating by gravity waves aid the initiation process of sprites and hence explain the association of sprites with MILs.

Further, observations of both MIL and lightning without sprites have been explained by considering the generation of gravity waves during deep convective activity and its propagation to mesosphere where they may help in the production of MIL and modulation of neutral density. However, convective systems with weak $\Delta M_q$ will not be able to initiate sprite and thus one can explain the observation of MIL and lightning without sprite.

The observation of lightning without MIL and sprite is easily explained by considering the fact that gravity waves are not able to reach the mesosphere. This situation may arise either due to absorption/filtering effect of gravity wave by the ambient medium or by the non-generation of gravity wave during thunderstorm activity. Further, case-by-case study in required.

**Acknowledgements**

The authors thank to Dr Torsten Neubert and Dr O. Chanrion for providing sprite and lightning data. One of the authors (Devendraa Siingh) wishes to thank Dr Torsten Neubert for general discussion on Sprites and MIL during his visit to the Danish National Space Center (DNSC), Copenhagen, Denmark. DS expresses his gratitude to Dr Jens Olaf P Pedersen for extending an invitation to visit DNSC. SF and DS acknowledge the Ministry of Earth Sciences, Government of India for financial support and also thanks to Dr. B. N. Goswamy, Director IITM, for his encouragement.

The authors thank anonymous reviewers (I & III reviewer) for their critical comments and valuable suggestions, which helped in improving the scientific value of this research paper. They also express their gratitude to Dr. E. R. Williams (II reviewer, MIT, USA) for a comprehensive and very thorough review, which resulted in significant improvements in the manuscript.

**Legends**

Figure 1:   Diagram showing the standard vertical thermal structure of the atmosphere, a temperature profile obtained from the SABER instrument aboard the TIMED satellite (blue profile) showing a MIL, where a MIL occurs above the thunderstorm (modified figure after Lyons et al., 2000; Neubert, 2003).

Figure 2:   Vertical temperature structure for the randomly selected typical inversion days (a) 22-07-2003 (near Pic du Midi), (b) 27-07-2003 (Puy de Dôme), (c) 11-08-2005 (near Puy de Dôme), (d) 09-09-2005 (near Puy de Dôme), (e) 27-07-2006 (near Puy de Dôme),(f) 13-10-2007 (near Pic du Midi) as recorded by SABER, where simultaneously sprites are observed above stratiform region of convective thunderstorms on the same day.

Figure 3:   Vertical temperature structure for the randomly selected typical non-inversion days (a) 25-07-2003 (near Pic du Midi), (b) 26-08-2003 (near Puy de Dôme), (c) 18-07-2005 (near Puy de Dôme), (d) 20-07-2005 (near Puy de Dôme), (e) 06-07-2006 (near Puy de Dôme), (f) 14-10-2007 (near Pic) as recorded by SABER, where simultaneously sprites are not observed on the same day.

Figure 4:   The variation of lightning flashes per ten minutes with time recorded at the Pic du Midi ($42^0\ 56'\ 11''$N; $00^0\ 08'\ 34''$E; 2.877 km altitude) in the French Pyrénées on August 13, 2005 and July 01, 2006.

Figure 5:   Vertical temperature structure for the randomly selected typical inversion days, which were not associated with sprites; (a) 31-07-2003, (b) 09-08-2003, (c) 08-07-2005, (d) 25-10-2005, (e) 23-07-2006 and (f) 23-08-2006

Table 1: Days (year-month-day) on which lightning are observed but no evidence of sprites at Pic du Midi (42⁰ 56' 11''N; 00⁰ 08' 34''E; 2.877 km altitude) in the French Pyrénées

| Day of observation | Occurrence of lightning | Range of times, lightning were observed (but no sprite), UT hr: min | Occurrence of Sprite |
|---|---|---|---|
| 2003-07-25 | Yes | 11:44 – 18:03 | No |
| 2003-07-26 | Yes | 01:56 – 23:59 | No |
| 2003-07-29 | Yes | 00:00 – 23:15 | No |
| 2003-08-01 | Yes | 01:56 – 23:59 | No |
| 2003-08-16 | Yes | 00:00 – 23:59 | No |
| 2003-08-26 | Yes | 00:00 -18:07 | No |
| 2005-07-12 | Yes | 00:40 – 23:59 | No |
| 2005-07-15 | Yes | 06:29 – 23:59 | No |
| 2005-08-13 | Yes | 11:22 – 22:38 | No |
| 2005-08-17 | Yes | 00:45 – 23:59 | No |
| 2005-09-06 | Yes | 00:00 – 23:59 | No |
| 2005-09-19 | Yes | 00:00 – 23:59 | No |
| 2005-10-30 | Yes | 06:19 – 23:59 | No |
| 2006-06-26 | Yes | 00:00 – 23:59 | No |
| 2006-07-01 | Yes | 00:00 – 23:59 | No |
| 2006-07-10 | Yes | 14:41- 22:24 | No |
| 2006-08-06 | Yes | 00:00 – 21:12 | No |
| 2006-08-15 | Yes | 00:00 – 23:59 | No |
| 2007-07-30 | Yes | 00:00 – 23:59 | No |
| 2007-08-05 | Yes | 00:00 – 23:59 | No |
| 2007-08-10 | Yes | 00:00 – 23:59 | No |
| 2007-08-11 | Yes | 00:00 – 23:59 | No |
| 2007-08-30 | Yes | 00:00 – 23:59 | No |
| 2007-09-01 | Yes | 00:00 – 23:24 | No |
| 2007-09-15 | Yes | 00:00 – 22:08 | No |
| 2007-09-30 | Yes | 00:00 – 23:59 | No |

Table 2: Days (year-month-day) on which lightning and sprites are observed at Pic du Midi (42⁰ 56' 11''N; 00⁰ 08' 34''E; 2.877 km altitude) in the French Pyrénées

| Day of observation | Occurrence of lightning | Range of times, lightning were observed, UT hr:min | Occurrence of Sprite | Range of times, sprites were observed, UT hr:min |
|---|---|---|---|---|
| 2003-07-21 | Yes | 00:00 – 23:59 | Yes | 02:05 - 03:15 |
| 2003-07-22 | Yes | 00:00 – 23:59 | Yes | 21:50 - 21:57 |
| 2003-07-23 | Yes | 00:00 – 23:59 | Yes | 21:11 – 23:21 |
| 2003-07-24 | Yes | 00:00 – 23:59 | Yes | 00:23 - 00:33 |
| 2005-07-29 | Yes | 00:00 – 23:59 | Yes | 01:28 – 02:04 |
| 2005-08-11 | Yes | 00:00 – 23:59 | Yes | 22:44 – 23:12 |
| 2005-09-09 | Yes | 00:00 – 23:59 | Yes | 01:81 – 21:18 |
| 2005-09-26 | Yes | 00:00 – 23:59 | Yes | 01:06 – 03:38 |
| 2006-07-04 | Yes | 00:00 – 23:59 | Yes | 20:46 – 22:51 |
| 2006-07-05 | Yes | 00:00 – 23:59 | Yes | 00:04 – 03:05 |
| 2006-07-22 | Yes | 00:00 – 23:59 | Yes | 01:46 – 02:01 |
| 2006-07-27 | Yes | 00:00 – 23:59 | Yes | 01:13 – 02:36 |
| 2006-08-17 | Yes | 00:00 – 23:59 | Yes | 22:29 – 23:57 |
| 2006-08-18 | Yes | 00:00 – 23:59 | Yes | 00:14 |
| 2007-08-29 | Yes | 00:00 – 23:59 | Yes | 20:11 – 20:37 |
| 2007-09-17 | Yes | 00:00 – 23:59 | Yes | 21:25 – 22:01 |
| 2007-10-13 | Yes | 00:00 – 23:59 | Yes | 00:05 – 03:54 |
| 2007-10-18 | yes | 00:00 – 23:59 | yes | 03:00 |

Table 3: Inversion days (year-month-day) observed in the SABER temperature profile and at least one sprite reported in the vicinity ($3^0 \times 3^0$ grid) on the same day (* indicate the plots of the given date)

| Date | Existence of sprite | | Evidence of MIL in SABER temperature profile at the location | | Range of times, sprites were observed, UT | Time of SABER temperature profile (UT) |
|---|---|---|---|---|---|---|
| | Lat. degrees N | Long. degrees E | Lat. degrees N | Long. degrees E | hr:min | hr:min:sec |
| 2003-07-21 | 46.8 | 4.1 | 46.2 | 5.2 | 02:05 - 03:15 | 03:33:26 |
| 2003-07-22 * | 45.0 | 5.6 | 42.3 | 3.0 | 21:50 - 21:57 | 07:16:23 |
| 2003-07-23 | 43.9 | 5.9 | 41.6 | 6.4 | 21:11 – 23:21 | 02:28:23 |
| 2003-07-24 | 44.6 | 6.8 | 46.1 | 7.6 | 00:23 - 00:33 | 02:47:56 |
| 2003-07-27 * | 46.5 | 9.7 | 44.3 | 6.2 | 20:46 - 21:13 | 02:01:25 |
| 2003-08-20 | 45.1 | 6.8 | 47.3 | 8.5 | 20:19 | 01:00:19 |
| 2003-08-25 | 42.7 | 10.8 | 40.1 | 10.2 | 19:56 – 21:37 | 00:52:52 |
| 2003-08-28 | 45.5 | 5.6 | 43.0 | 9.0 | 20:18 - 23:34 | 19:17:24 |
| 2003-08-29 | 46.0 | 6.0 | 48.0 | 4.0 | 00:04 – 02:35 | 22:59:46 |
| 2005-07-19 | 45.3 | 9.9 | 47.9 | 10.4 | 00:46 | 03:37:18 |
| 2005-07-29 | 45.2 | 2.6 | 45.0 | 5.5 | 01:28 – 02:04 | 01:29:03 |
| 2005-08-11 * | 43.5 | 6.7 | 42.0 | 2.0 | 22:44 – 23:12 | 03:40:13 |
| 2005-08-16 | 41.9 | 0.8 | 48.4 | 1.7 | 23:06 | 22:38:43 |
| 2005-09-09 * | 43.84 | 5.46 | 43.7 | 6.1 | 01:81 – 21:18 | 21:19:32 |
| 2005-10-23 | 40.6 | 3.7 | 41.1 | 8.8 | 18:59 | 17:19:03 |
| 2005-11-11 | 40.5 | 3.3 | 40.6 | 3.6 | 00:56 – 02:56 | 23:40:29 |
| 2006-07-04 | 44.3 | -0.808 | 41.3 | 5.4 | 20:46 – 22:51 | 02:03:21 |
| 2006-07-27 * | 45.9 | 1.0 | 11.1 | 0.9 | 01:13 – 02:36 | 11:57:36 |
| 2006-08-18 | 44.7 | 5.9 | 41.9 | 5.8 | 00:14 | 20:52:26 |
| 2007-08-29 | 46.2 | 6.9 | 46.3 | 9.8 | 20:11 – 20:37 | 22:40:25 |
| 2007-09-17 | 44.6 | 5.4 | 41.8 | 9.5 | 21:25 – 22:01 | 14:38:13 |
| 2007-10-13 * | 37.7 | 5.8 | 38.7 | 3.7 | 00:05 – 03:54 | 05:22:02 |
| 2007-10-18 | 39.44 | 1.1 | 36.2 | 2.7 | 03:00 | 18:32:26 |

Table 4: Non inversion days (year-month-day) observed in the SABER temperature profile and no sprite reported in the vicinity ($3^0 \times 3^0$ grid) on the same day (* indicate the plots of the given date)

| Day of observation | No evidence of sprite at the location | | No evidence of MIL in SABER temperature profile at the location | | Range of times, lightning were observed (but no sprite), UT | Time of SABER temperature profile (UT) |
|---|---|---|---|---|---|---|
| | Lat. degrees N | Long. degrees E | Lat. degrees N | Long. degrees E | hr:min | hr:min:sec |
| 2003-07-25 * | 44.0 | 5.8 | 48.7 | 8.0 | 11:44 – 18:03 | 03:06:56 |
| 2003-07-29 | 47.1 | 9.5 | 48.4 | 2.2 | 00:00 – 23:15 | 02:38:25 |
| 2003-07-30 | 47.1 | 9.5 | 45.0 | 10.0 | 00:29 – 23:59 | 01:16:20 |
| 2003-08-21 | 45.1 | 6.8 | 45.9 | 4.7 | 00:00 – 23:59 | 01:18:22 |
| 2003-08-26 * | 42.0 | 11.0 | 44.0 | 13.8 | 00:00 – 18:07 | 23:48:14 |
| 2003-08-27 | 46.9 | 10.4 | 46.0 | 9.8 | 00:24 – 23:59 | 18:58:49 |
| 2003-08-30 | 46.9 | 10.4 | 44.7 | 7.8 | 00:00 – 23:59 | 23:19:29 |
| 2003-08-31 | 46.9 | 10.4 | 49.6 | 10.6 | 00:26 – 23:59 | 21:56:11 |
| 2005-07-18 * | 45.3 | 9.9 | 45.2 | 10.6 | 00:00 – 23:59 | 03:19:09 |
| 2005-07-20 * | 45.3 | 9.9 | 48.9 | 6.7 | 14:02 – 17:49 | 03:54:35 |
| 2005-07-28 | 45.3 | 9.9 | 43.0 | 8.6 | 00:00 – 23:59 | 01:11:30 |
| 2005-07-30 | 46.2 | 3.6 | 47.6 | 4.7 | 00:00 – 23:59 | 01:47:05 |
| 2005-08-09 | 43.9 | 6.8 | 45.0 | 10.8 | 00:00 – 23:59 | 03:05:45 |
| 2005-08-10 | 43.9 | 6.8 | 42.8 | 7.9 | 00:00 – 23:59 | 03:23:23 |
| 2005-08-12 | 43.9 | 6.8 | 47.8 | 4.0 | 00:00 - 18:59 | 23:09:29 |
| 2005-08-13 | 43.8 | 6.8 | 42.8 | 10.9 | 11:22 - 23:59 | 02:35:40 |
| 2005-08-14 | 43.9 | 6.8 | 42.4 | 1.5 | 00:00 – 23:59 | 22:02:08 |
| 2005-08-15 | 43.9 | 6.8 | 48.3 | 8.2 | 00:00 – 23:59 | 01:28:17 |
| 2005-08-17 | 43.8 | 7.0 | 42.6 | 7.3 | 00:45 - 23:59 | 02:04:34 |
| 2005-09-05 | 43.0 | 6.0 | 46.5 | 3.9 | 00:00 – 23:59 | 21:49:12 |
| 2005-09-06 | 41.0 | 0.8 | 37.2 | 2.8 | 00:00 – 23:59 | 16:59:05 |
| 2005-09-08 | 43.0 | 5.4 | 46.4 | 7.4 | 00:00 – 23:59 | 21:01:35 |
| 2005-09-15 | 43.9 | 6.8 | 46.2 | 2.3 | 02:40 - 23:59 | 22:20:45 |
| 2005-09-25 | 43.0 | 3.2 | 41.5 | 2.0 | 04:01 – 23:59 | 09:11:04 |
| 2005-09-26 | 40.12 | 3.8 | 40.8 | 4.8 | 00:00 – 23:59 | 22:52:00 |
| 2006-07-06 * | 45.1 | 2.6 | 43.0 | 4.8 | 00:00 – 23:59 | 00:59:24 |
| 2006-08-19 | 45.5 | 5.1 | 45.2 | 4.7 | 00:00 – 23:59 | 21:10:25 |
| 2007-10-12 | 36.1 | 9.9 | 38.8 | 10.7 | 00:00 – 23:59 | 05:05:53 |
| 2007-10-14 * | 38.0 | 2.6 | 38.0 | 6.2 | 00:00 – 22:11 | 19:04:01 |

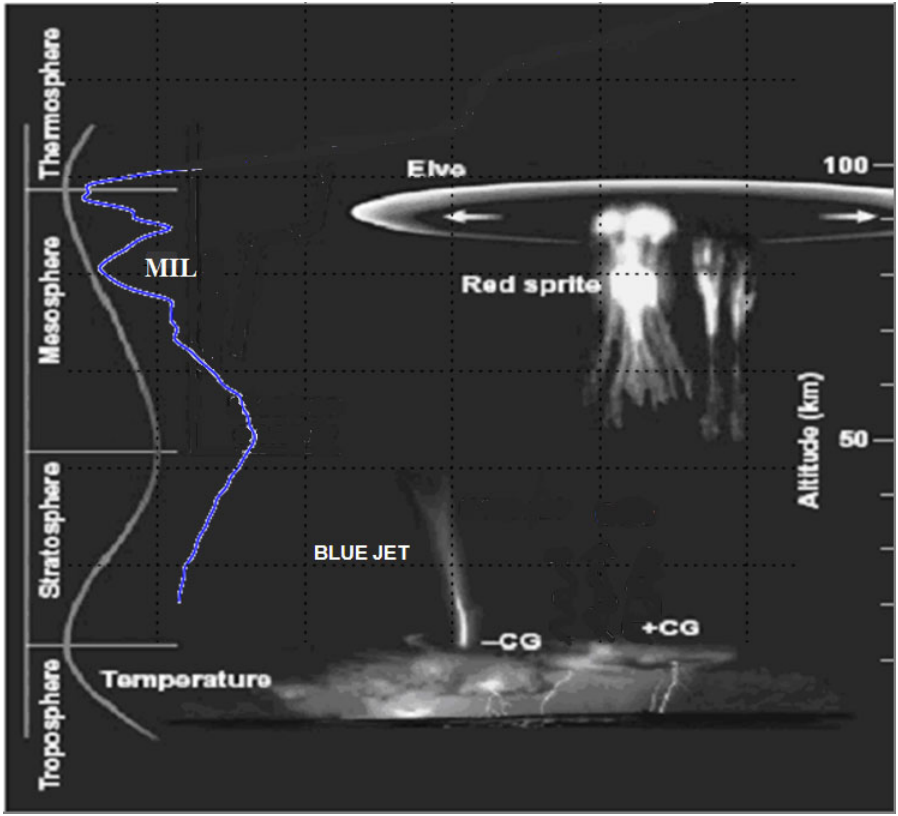

Figure.1

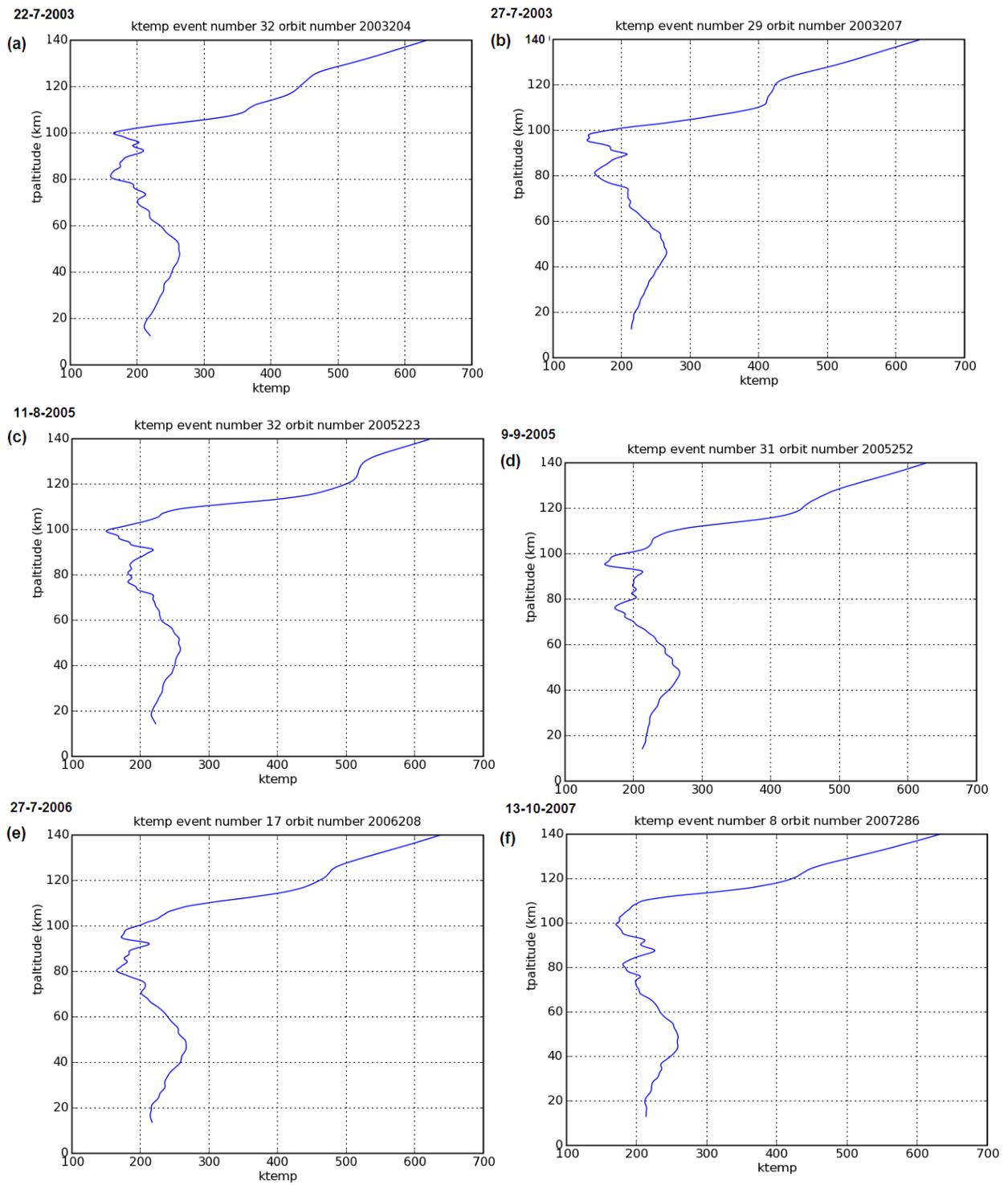

Figure 2

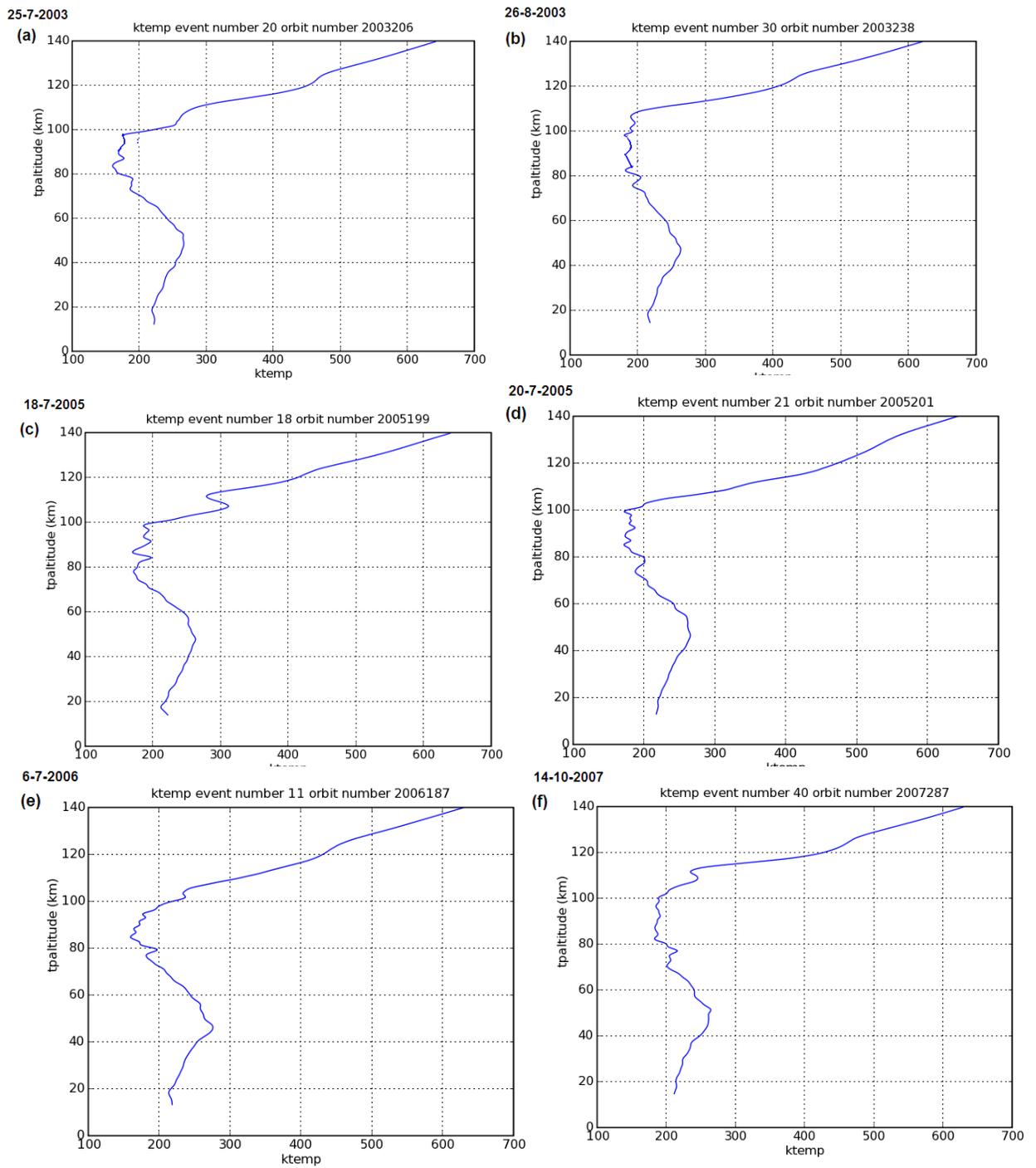

Figure 3

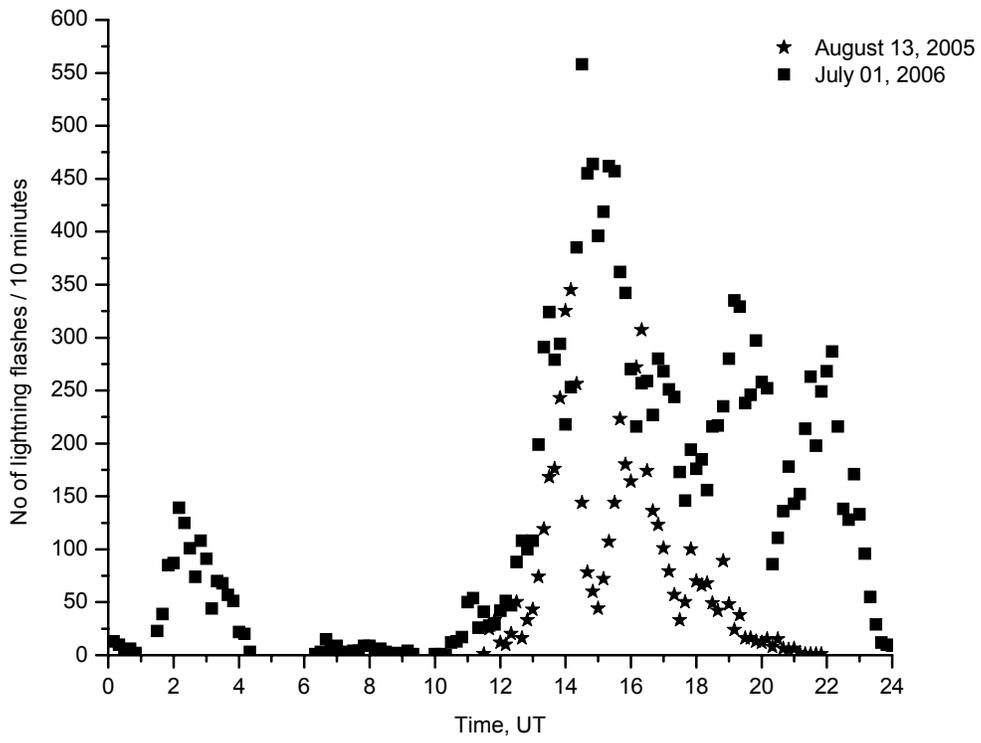

Figure 4

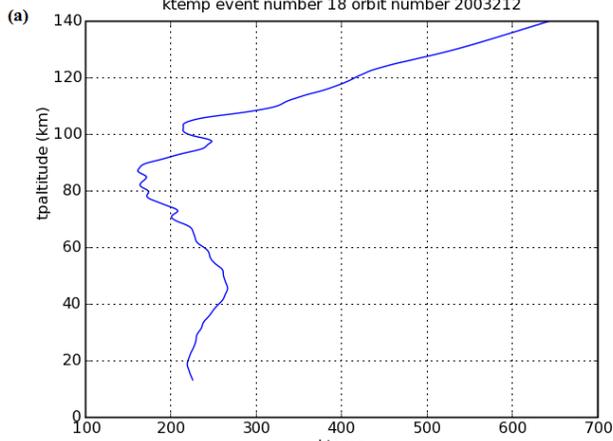
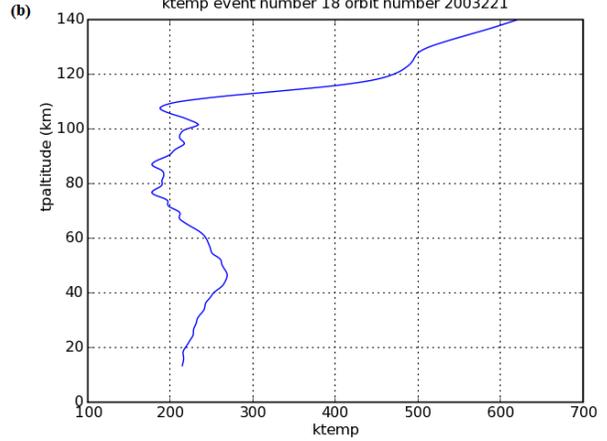
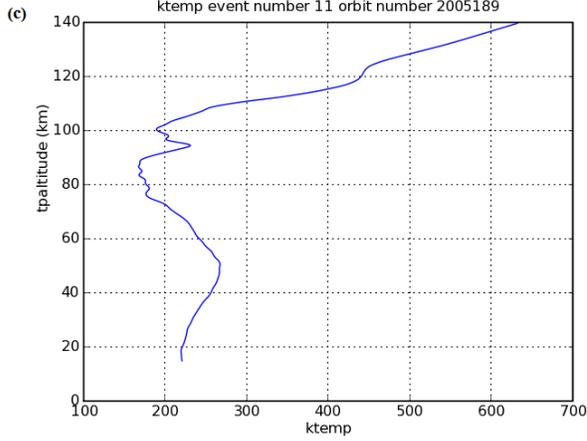
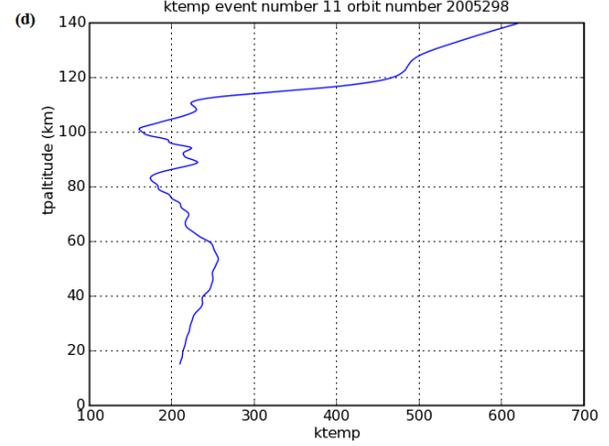
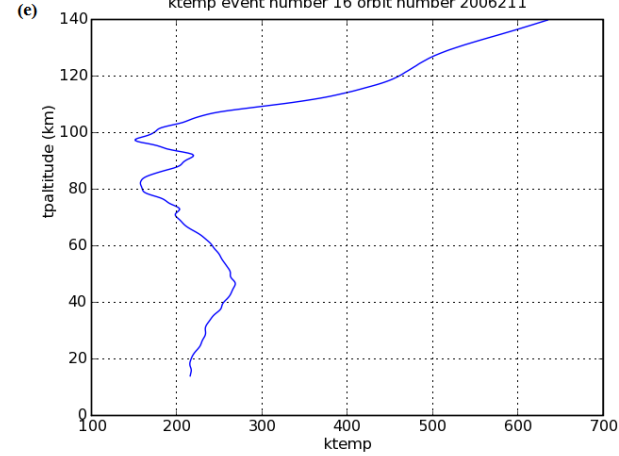
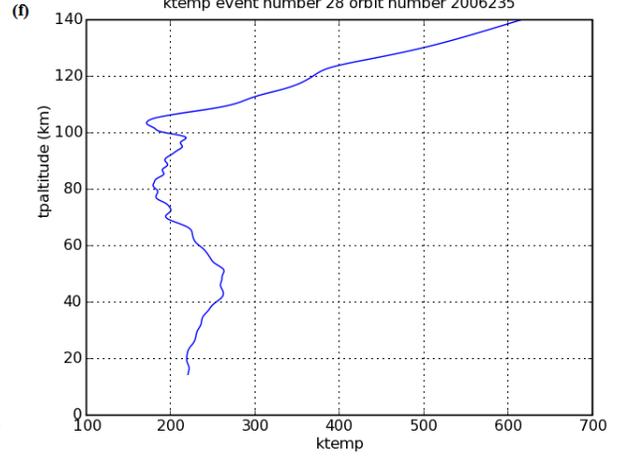

Figure 5

34